\newcommand{\supr}[1]{\ensuremath{^{\textrm{#1}}}}
\newcommand{\subs}[1]{\ensuremath{_{\textnormal{#1}}}}

\documentclass[prb,aps,reprint]{revtex4-1}
\usepackage{graphicx}
\begin{document}

\title{First-principles study of the energetics of charge and cation mixing in U$_{\rm 1-x}$Ce$_{\rm x}$O$_{\rm 2}$}
\author{B. E. \surname{Hanken}}
\affiliation{Department of Chemical Engineering and Materials Science, University of California, Davis CA 95616 USA}
\author{C. R. \surname{Stanek}}
\affiliation{Materials Science \& Technology Division, Los Alamos National Laboratory, Los Alamos, NM 87545 USA}
\author{N. \surname{Gr{\o}nbech-Jensen}}
\affiliation{Department of Applied Science, University of California, Davis CA 95616 USA}
\author{M. Asta}
\affiliation{Department of Materials Science and Engineering, University of California, Berkeley CA 94720 USA}
\affiliation{Department of Chemical Engineering and Materials Science, University of California, Davis CA 95616 USA}
\email[Contact author at ]{mdasta@berkeley.edu}

\date{\today}

\begin{abstract}
The formalism of electronic density-functional-theory, with Hubbard-U corrections (DFT+U), is employed in a computational study of the energetics of U$_{\rm 1-x}$Ce$_{\rm x}$O$_{\rm 2}$ mixtures. The computational approach makes use of a procedure which facilitates convergence of the calculations to multiple self-consistent DFT+U solutions for a given cation arrangement, corresponding to different charge states for the U and Ce ions in several prototypical cation arrangements. Results indicate a significant dependence of the structural and energetic properties on the nature of both charge and cation ordering. With the effective Hubbard-U parameters that reproduce well the measured oxidation-reduction energies for urania and ceria, we find that charge transfer between U(IV) and Ce(IV) ions, leading to the formation of U(V) and Ce(III), gives rise to an increase in the mixing energy in the range of 4-14 kJ/mol of formula unit, depending on the nature of the cation ordering. The results suggest that although charge transfer between uranium and cerium ions is disfavored energetically, it is likely to be entropically stabilized at the high temperatures relevant to the processing and service of urania-based solid solutions.
\end{abstract}

\maketitle

\section{Introduction}
The thermophysical properties of urania-ceria mixtures have been extensively researched for more than 30 years \cite{Kim06,Lindemer86,Markin70}.  Interest in this system stems primarily from two issues related to the performance of oxide nuclear fuel.  First, cerium is a fission product in urania nuclear fuels that is highly soluble in the fluorite UO\subs{2} structure \cite{Markin70}; its effects on phase stability, thermal conductivity and ionic diffusion are thus important issues relevant to nuclear fuel performance, particularly for future potential applications involving high burnup.  Second, urania-ceria has been investigated as a surrogate system for urania-plutonia mixed-oxide (MOX) fuels, owing to the similar oxidation states and ionic radii for Ce and Pu ions.  In spite of the interest in this system, and the extensive experimental research that has been devoted to it to date, the thermodynamic properties and phase diagram of urania-ceria remain incompletely characterized.

Due to the importance of the solution-thermodynamic properties and phase equilibria on the performance of nuclear fuels, renewed efforts have been undertaken to develop accurate thermochemical models for these materials.  A prominent approach is the application of the so-called CALPHAD (Calculation of Alloy Phase Diagrams) methodology \cite{Lukas07}, involving the parametrization of compound-energy-formalism \cite{Hillert01} free-energy models for UO\subs{2} and its mixtures \cite{Gueneau02,Yamanaka04}.  For these efforts, a challenging issue is the lack of complete databases of experimental thermodynamic data, as well as the often conflicting nature of the results that are available.  In CALPHAD modeling of metal alloy systems, a strategy that has been effectively pursued to improve the accuracy of the thermodynamical models involves the application of first-principles calculations as a framework to augment available experimental data in the development of robust thermodynamic databases \cite{Ghosh02,Ghosh07,Turchi07,Liu09}. Over the last decade advances in first-principles methods have led to widespread use of these techniques in the modeling of defect structures and energetic properties in pure UO\subs{2} \cite{Nerikar09,Andersson09,Geng08}.  To date, however, they have not been applied extensively in the study of mixed oxide thermodynamic properties.

The present work involves application of first-principles methods, based on the Density Functional Theory with Hubbard-U (DFT+U) formalism (see section \ref{ss:calcs}), to the study of mixing energetics in urania-ceria solid solutions with stoichiometric oxygen compositions.  The work represents an initial step towards more complete modeling of solution thermodynamics for both stoichiometric and non-stoichiometric oxygen compositions in this system.  It also provides insight into an issue that has been extensively discussed in the context of the properties of urania-ceria solid solutions, namely the presence of mixed charge states for the uranium and cerium cations. Specifically, magnetic susceptibility \cite{Hinatsu88} and X-Ray Photoemission Spectroscopy (XPS) measurements have shown evidence of charge-transfer processes by which a fraction of the U(IV) and Ce(IV) ions are oxidized and reduced, respectively, to U(V) and Ce(III) \cite{Bera09}.  However, conflicting conclusions have been drawn from X-Ray Adsorption Near-Edge Spectra (XANES) measurements \cite{Antonio96} where it was concluded that Ce and U retain oxidation states of IV in Ce-rich solid solutions. The current work presents a computational methodology that can be used to systematically investigate the relative energetics of solid solutions with ideal versus mixed-charge states, and suggests a picture in which U$_{\rm 1-x}$Ce$_{\rm x}$O$_{\rm 2}$ compounds with mixed charge states are energetically disfavored while being entropically stabilized at high temperatures.

The remainder of this paper is organized as follows: we begin by describing the computational methodology and follow with an explanation of how we converge to and characterize different charge states in the DFT+U calculations. The following section presents results for energetics, ionic structure, and electronic structure, augmented with classical-pair-potential studies extending the work to disordered systems. We conclude by discussing implications of the first-principles results for the mixing thermodynamics in this system.

\section{Methods}
\begin{figure}
\includegraphics[width=\columnwidth]{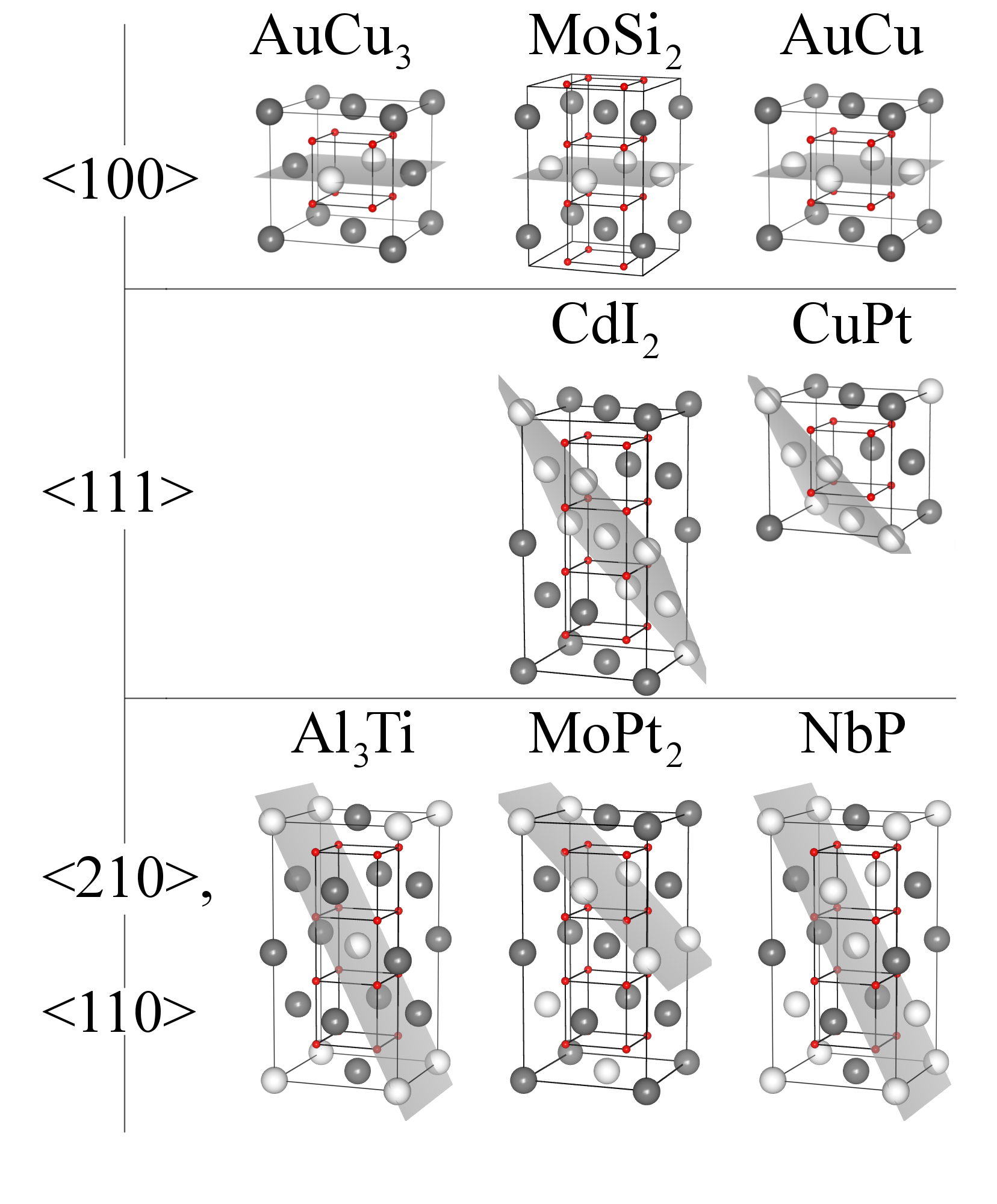}
\caption{(color online) An illustration of the prototype ordered structures considered in the DFT+U calculations for the five compositions of mixed oxide explored.  Ordering waves are identified, and the stacking planes are shown in gray. The oxygen cubic sublattice is also shown.}
\label{fig:prototyp}
\end{figure}

To investigate the energetics of charge and cation ordering in the urania-ceria system, we consider a set of fluorite-based superstructures in which the cations are arranged in accordance with several fcc-based prototypical ordered compounds. Cation arrangements in these structures are built from ordering waves along the $<$100$>$, $<$210$>$ and $<$111$>$ ``special-point" ordering directions \cite{Ducastelle1991}, and include the so-called ``Lifshitz structures" for ordering on the fcc cation sublattice of the fluorite structure.  All structures considered in this work are stoichiometric, i.e., the oxygen fluorite sublattice is fully occupied with no interstitial oxygen atoms present.  Experimentally, urania-ceria solid solutions are observed to be configurationally disordered on the cation sublattice.  However, in the present work we consider only the ordered structures shown in Fig. \ref{fig:prototyp} and listed in Table \ref{tab:structs} in the DFT+U calculations.  To model the energetics of disordered solid solutions, we employ pair-potential models tuned to the DFT+U results, in supercell calculations of random mixtures.

\begin{table}
\begin{tabular}{c|c|c|c}
Composition & Prototype & Strukturbericht & Ordering Wave \\
\hline
A\subs{3}B & AuCu\subs{3} & L1\subs{2} & $<$100$>$ \\
A\subs{2}B & MoSi\subs{2} & C11\subs{b} & \\
A\subs{2}B\subs{2} & AuCu  & L1\subs{0} &  \\
\hline
A\subs{2}B & CdI\subs{2} & C6 & $<$111$>$ \\
AB & CuPt & L1\subs{1} & \\
\hline
A\subs{3}B & Al\subs{3}Ti & D0\subs{22} & $<$210$>$\\
A\subs{2}B & MoPt\subs{2} & & \\
A\subs{4}B\subs{4} & NbP & & \\
\end{tabular}
\caption{Compositions, prototypes, Strukterbericht and associated ordering wave families for the prototype ordered structures considered in the DFT+U calculations.}
\label{tab:structs}
\end{table}

In our calculations we consider two types of ionic charge states.  In the first, which we will refer to as ``ideal charge states" (ICS), all cerium and uranium ions possess the 4+ charge state (see discussion below for the definition of the charge states based on the results of the electronic structure calculations).  In the second, which we will refer to as ``mixed charge states" (MCS), a fraction of the cerium and uranium ions are respectively reduced and oxidized to 3+ and 5+ charge states in a 1:1 ratio of Ce(III):U(V) to preserve overall charge neutrality.

\subsection{First Principles Calculations}
\label{ss:calcs}
All DFT+U calculations have been performed employing the formalism of Dudarev {\em et al.} \cite{Dudarev98}, as implemented in the Vienna ab-initio simulation package (VASP) \cite{Kresse93,Kresse96a,Kresse96b}.  In the implementation used in this work, we have employed the projector augmented-wave (PAW) method \cite{Blochl94,Kresse99} and the Perdew-Becke-Ernzerhof generalized gradient approximation (PBE-GGA) \cite{Perdew96,Perdew97}.  The PAW-PBE potentials employed in this work are those designated ``U", ``Ce", and ``O" in the VASP library.  The electronic wave functions were expanded in a plane-wave basis set with a cutoff of 500 eV, and the electronic states were sampled using k-point meshes centered on the origin, with a density equivalent to that of a 4x4x4 mesh in the Brillouin zone of the ideal fluorite structure.  The structures were fully relaxed with no symmetry constraints imposed, until the magnitude of the forces was below 10\supr{-3} eV/\AA\ and the pressure was below 1 kBar.  With these settings the calculated mixing energies are estimated to be converged to within 1 meV per (U,Ce)O\subs{2} formula unit (FU). The mixing enthalpy is defined as follows:
\begin{equation}
H_{mix} = H_{U_{x}Ce_{1-x}O_{2}} - ( x H_{UO_{2}} +  ( 1 - x ) H_{CeO_{2}} ),
\end{equation}
where $H_{U_xCe_{1-x}O_2}$ denotes the energy (per cation) of a urania-ceria mixture with $x$ denoting the cation mole fraction of uranium ions, and similarly for $H_{UO_2}$ and $H_{CeO_2}$. All of the results presented in this manuscript were calculated using the scalar-relativistic approximation. For the L1\subs{0} ICS and MCS structures, we also performed calculations including spin-orbit coupling and the resulting mixing energies were found to change by a magnitude of less than 3.5 meV/FU for the ICS, and 25 meV/FU for the MCS.

The differences in energy between MCS and ICS for a given structure are highly sensitive to the choice of the parameter U\subs{eff}=U-J in the DFT+U formalism employed in this work.  The values of U\subs{eff} for U $5f$ and Ce $4f$ electrons utilized in the present calculations yield a close match between calculated and experimental values of relevant oxidation and reduction energies.  For uranium, the choice of U\subs{eff}=3.99 eV, which was proposed originally by Dudarev {\em et al.} \cite{Dudarev98} and has been widely used in the literature, was found in the present study to give rise to good agreement between calculated and measured values of the enthalpy changes for the following two oxidation reactions:  (i) UO\subs{2} + $\frac{1}{2}$ O\subs{2} $\rightarrow$ $\gamma$-UO\subs{3}, (ii) 3 UO\subs{2} + O\subs{2} $\rightarrow$ $\alpha$-U\subs{3}O\subs{8}.  Descriptions of $\gamma$-UO\subs{3} and the $\alpha$-U\subs{3}O\subs{8} can be found in Refs. \cite{Engmann63} and \cite{Loopstra64}, respectively.  With U\subs{eff}=3.99 eV, and employing the correction for over-binding of the oxygen molecule discussed in Ref. \cite{Wang06}, we obtain values of $\Delta$ H\subs{rxn}=-1.45 eV and -3.44 eV for reactions (i) and (ii), which agrees well with the corresponding experimental values of -1.44 eV and -3.31 eV, as determined from the CODATA Key Values for Thermodynamics \cite{Codata}.  For cerium we choose the value U\subs{eff}=3 eV, which was shown by Andersson {\em et al.} \cite{Andersson07} to yield good agreement between calculated and experimental values for the enthalpy of the reduction reaction:  Ce\subs{2}O\subs{3} + $\frac{1}{2}$ O\subs{2} $\rightarrow$ 2 CeO\subs{2}.

\subsection{Electronic and Ionic Relaxation}
\label{ss:appr}
An important issue that has been recently discussed in the literature \cite{Dorado09,Zhou09,Meredig10} concerns the propensity for DFT+U calculations to converge to multiple self-consistent solutions corresponding to different orbital occupations.  To ensure that such calculations converge to solutions that are the lowest-energy electronic states (or near them), several different methods have been proposed.  In the current work we employ the approach described by Meredig {\em et al.}\cite{Meredig10}, which involves a slow localization of the $f$ electrons in a series of DFT+U calculations with incrementally increasing values of U\subs{eff}.  Specifically, one begins by performing a GGA (with U\subs{eff}=0) calculation until ionic positions and charge density are converged.  The wavefunction eigenvalues, charge densities, and atomic positions are then used as the starting point for a calculation with a small non-zero value of U\subs{eff}, again performed to convergence of ionic position and charge density.  The results of this calculation are then used as the starting point for a calculation with an incrementally higher U\subs{eff}, and this process is continued until the desired U\subs{eff} is reached. Steps in U\subs{eff} of approximately 0.1 eV have been found sufficient to reproducibly converge the low energy structure of UO\subs{2}.

For urania-ceria mixtures, this method for converging the electronic structure becomes somewhat ambiguous.  Specifically, U\subs{eff} can be incremented for each ion type simultaneously or in two separate stages.   However, this ambiguity has been exploited in the current work to enable convergence of the electronic structure to the ICS and MCS charge configurations.  Indeed, we found that the ICS charge state (i.e., composed of Ce(IV) and U(IV)) is obtained if U\subs{eff} is first ramped up to its final value on uranium $5f$ electrons, while holding the value of U\subs{eff}=0 for the cerium $4f$ electrons.  Alternatively, if U\subs{eff} is first applied to cerium $4f$ electrons, the MCS charge state (i.e., containing Ce(III) and U(V)) is obtained.

The oxidation states of the ions resulting from the calculations described above are identified based on three considerations:  the interatomic bond lengths, the number of states in the occupied electronic partial densities of states, and the local magnetic moments.  For structures with Ce(III) ions, the number of states in the occupied Ce $4f$ projected densities of states is found to integrate to approximately one electron (cf. Fig. \ref{fig:l10_pdos}), while this band lies above the Fermi level for structures containing only Ce(IV) ions.  Similarly, the number of states in the occupied U $5f$ projected densities of states is approximately two and one for U(IV) and U(V), respectively.  Corresponding to these different charge states are distinct values of the local magnetic moments.  For U(IV), U(V), Ce(III) and Ce(IV) ions the calculated local moments are close to the ideal values of 2, 1, 1 and 0, respectively, as expected from Hund's rule, which assumes that these moments originate from the single and paired $f$ electrons.  A final consistency check associated with the labeling of the charge states is made based on the calculated relaxed bond lengths.  Oxidation of U(IV) to U(V) is found to lead to the expected decrease in the average relaxed U-O bond lengths, while reduction of Ce(IV) to Ce(III) leads to an increase of the Ce-O bond lengths (cf. Fig. \ref{fig:bldist}).

\subsection{Classical Pair Potentials}
\label{ss:pp}

To understand the implications of the first-principles results for the thermodynamic properties of disordered solid solutions, we have used the energetic results presented in section \ref{s:res} to optimize ionic pair-potential models for urania-ceria solid solutions.  Specifically, we employ the polarizable shell-model potentials of the Buckingham type for O, U(IV), Ce(III) and Ce(IV) published in Refs. \cite{Binks93,Vyas98,Busker99,Levy04}.  A similar potential for U(V) has been previously developed by one of us \cite{Stanek}. The interaction parameters for this potential are A=2386.42 eV, $\rho$=0.3411 \AA\, and C=0 eV-\AA \supr{6}. For U(V), an important issue concerns the magnitude of the fifth ionization potential, which cannot be calibrated to any direct experimentally measurable properties, but is required in order to compare energetic differences between ICS and MCS states.  Calculations by Pyper and Grant estimate this value at 46.57 eV \cite{Pyper78}.  We found that an optimal agreement between potential-model predictions and the DFT+U energetics is obtained with a value of 47.62 eV, an increase of approximately 2\%.  Using this value for the fifth-ionization energy, and the published parametrizations for the other cations, the classical-potential models match the DFT+U mixing energies of x = 0.25, 0.50, and 0.75 in Fig. \ref{fig:energies} (discussed in more detail in the Results section) to within a standard deviation of 4.5 kJ/mol.

With the pair potentials, the focus on urania rich disordered solid solutions is motivated, in part, by ongoing experimental work in this system \cite{Hanken}. It is also within this regime that the solution thermodynamics of such materials is relevant to nuclear fuels in particular. The calculations employing ionic pair potentials proceeds as follows.  Supercells containing 256 cations were constructed with random occupations of the cation sublattice at compositions of 25, 21.9, 15.6, 12.5, 6.25, and 3.1 cation percent Ce. For each of these cation configurations, we also considered ICS states as well as two MCS states with 50 and 100 \% of the Ce ions reduced to the a charge state of III (and a corresponding number of U ions oxidized to a charge state of V to maintain charge neutrality). At each of the resultant 18 compositions, 300 supercells were generated with different random-number seeds for each composition.  For each of the constructed supercells, full geometry relaxations were performed employing the GULP software \cite{GULP}.

\section{Results}
\label{s:res}

\subsection{First-Principles Results}

\begin{figure}[h]
\includegraphics[width=\columnwidth]{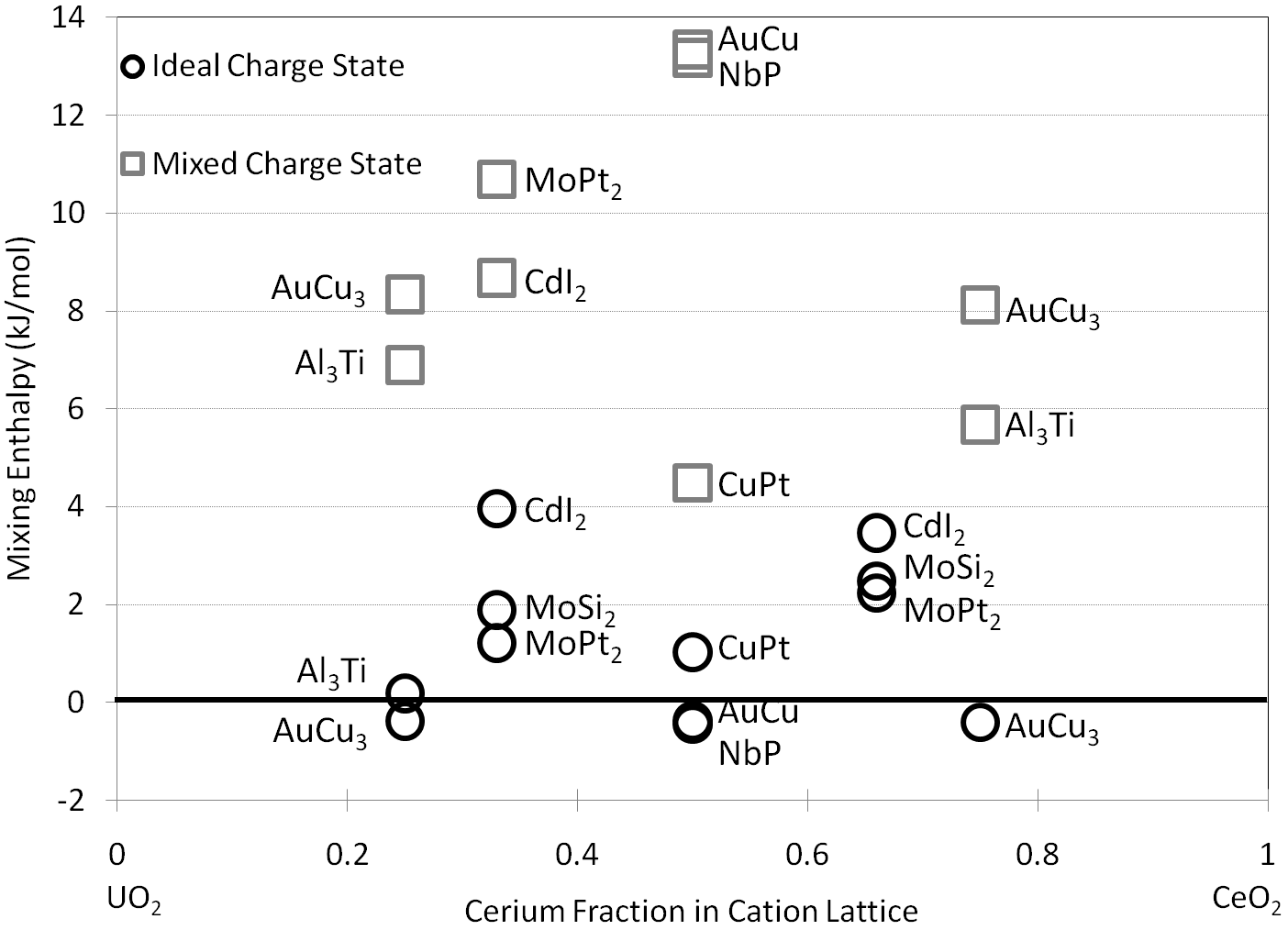}
\caption{Zero temperature mixing enthalpies for ordered structures as a function of Ce fraction on the cation sites. Energies for MCS compounds are shown with gray squares, while those for ICS compounds are indicated by black circles.}
\label{fig:energies}
\end{figure}

Figure \ref{fig:energies} plots DFT+U calculated values of the mixing energies for each of the prototype ordered structures shown in Fig. \ref{fig:prototyp}, for both ICS and MCS states. Considering first the ICS structures, the mixing energies are relatively small in magnitude, ranging from very weakly negative values of ~1 kJ/mol to positive values less than 5 kJ/mol in magnitude. The MCS structures are calculated to be higher in energy than their corresponding ICS analogues at the same composition. However, the difference in energy between ICS and MCS states is seen to be sensitive to the nature of the cation ordering. For example, focusing on the structures at equiatomic composition, the AuCu and NbP structures, which are built from $<$100$>$ and $<$1$\slantfrac{1}{2}$0$>$ special-point ordering waves, respectively, have energies that increase by 14 kJ/mol in going from the ICS to MCS states. By contrast, the CuPt structure, built from $<\slantfrac{1}{2}\slantfrac{1}{2}\slantfrac{1}{2}>$ ordering waves, is only 4 kJ/mol higher in energy in the ICS relative to the corresponding MCS state. A comparison of the results at each composition also shows an interesting trend that the structures with the highest energies in the ICS states have the lowest energies in the MCS states. In general, the variation in the energy differences between ICS and MCS states for the different structures suggests a significant coupling between the nature of the cation order and the energetics associated with charge transfer in the system.

\begin{figure}
\includegraphics[width=\columnwidth]{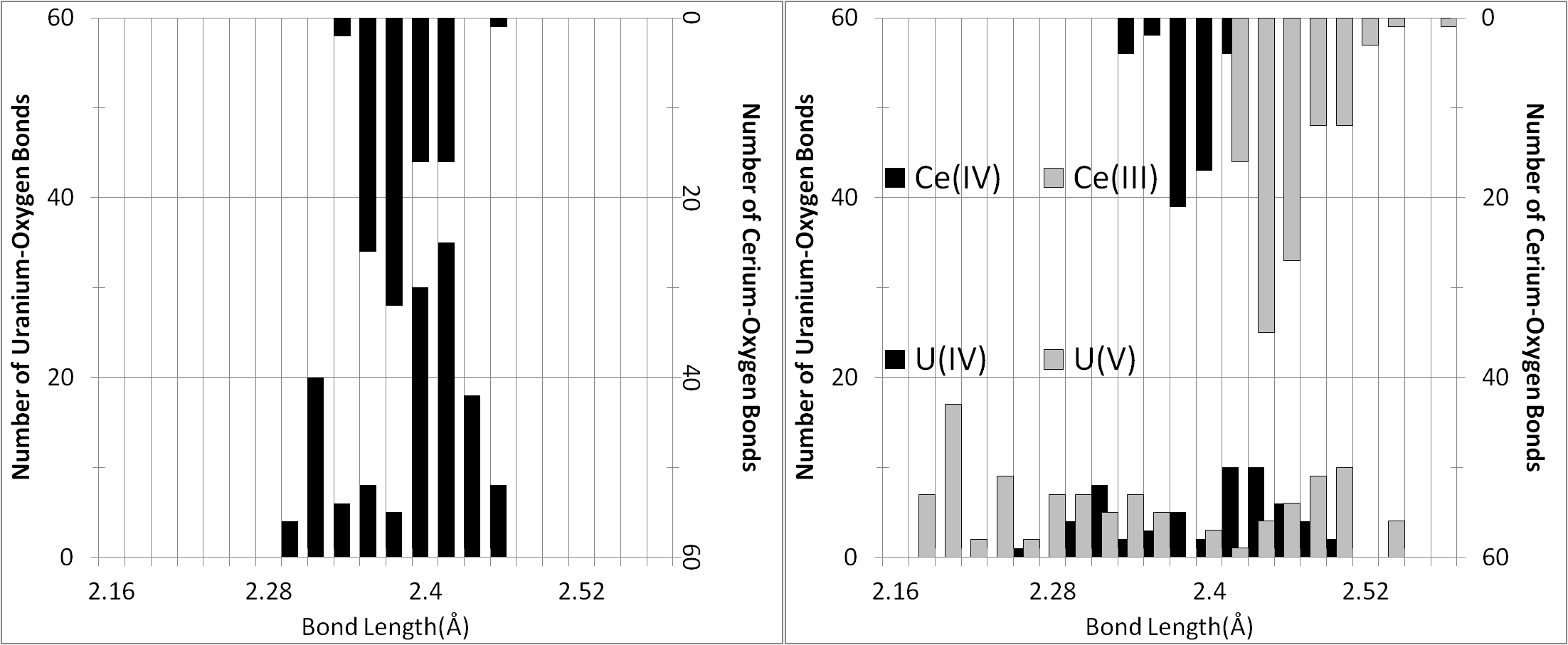}
\caption{Bond length distributions for ICS ordering (left panel) and MCS ordering (right panel).  Distributions are summed over all structures considered in the DFT+U calculations, amounting to a total of 384 metal-oxygen bonds for the ICS, and 352 for the MCS compounds.}
\label{fig:bldist}
\end{figure}

In Fig. \ref{fig:bldist} we compile results for the calculated atomic structures of the ICS and MCS compounds. The results are plotted in the form of distributions of bond lengths between the different cation species and their eight nearest-neighbor oxygen ions. These distributions are derived from the fully relaxed structures of all the ICS and MCS compounds considered in the DFT+U calculations. For ICS ordering, the averaged lengths for U(IV)-O and Ce(IV)-O bonds are 2.393 and 2.385 \AA, respectively, reflecting a slightly smaller size for the Ce(IV) ions relative to U(IV) in the fluorite structure. In the MCS compounds, the U(IV) and U(V) bond-length distributions overlap significantly, although the former show a clearly smaller average bond length, as expected: the averaged values for the U(IV)-O bond length in the MCS compounds is 2.399 \AA, which is similar to its averaged value in the ICS structures, while the corresponding value for U(V)-O in the MCS compounds is 2.336 \AA. The distribution of Ce-O bond lengths in the MCS compounds is seen to be significantly narrower than that of the U-O bonds in the same structures. The average Ce(IV)-O bond length in the MCS compounds has a value of 2.386 \AA\, which is very similar to that in the ICS structures. The Ce(III)-O bond length is larger, as expected, with an average value in the MCS compounds of 2.458 \AA. Overall, the results display the expected trends associated with changes in the cation radii with different oxidation states. The broader distributions for the U-O bond lengths in the MCS relative to the ICS compounds suggests a higher degree of strain energy in the former structures, with the preferred bond lengths for the different cation charge states being accommodated to differing degrees depending on the nature of the cation ordering.

\begin{figure}[h]
\includegraphics[width=\columnwidth]{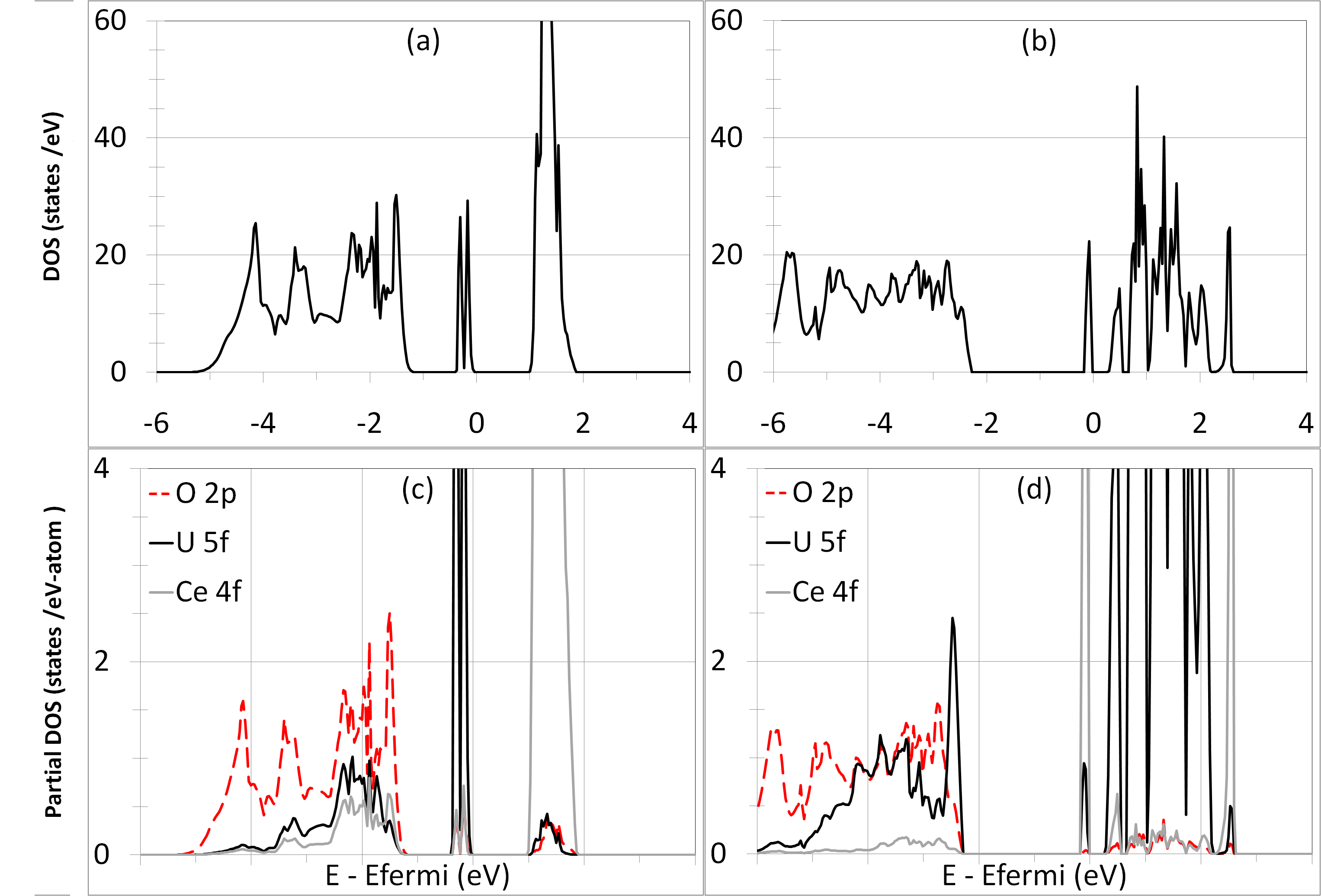}
\caption{(color online) Total density of states for AuCu ordered: a) ICS; and b) MCS. Partial density of states for AuCu ordered: c) ICS; and d) MCS.}
\label{fig:l10_pdos}
\end{figure}

The differences in the electronic structures for ICS versus MCS compounds is illustrated by the calculated electronic densities of states (DOS) for the equiatomic AuCu prototype structure in Fig. \ref{fig:l10_pdos}. In this figure, the ICS structure features a narrow band just below the Fermi level corresponding to the occupied 5$f$ orbitals for the U(IV) ions. This band is split off from the oxygen 2$p$ band with a gap of 0.80 eV. In the MCS compound, the U 5$f$ states shift to significantly lower energies with the highest density of states for these orbitals found at the edge of the O 2$p$ band.  The U(V)-O bonding thus displays a significantly stronger degree of covalent character, characterized by a stronger hybridization between the U 5$f$ and O 2$p$ states in the MCS compound.  The occupied Ce 4$f$ states for the Ce(III) ion in the MCS compound are observed to be split off from the U 5$f$ and O 2$p$ bands by a gap of approximately 2 eV, and these states display a much lower degree of hybridization with the O 2$p$ orbitals.

\subsection{Pair-Potential Modeling}
\label{ss:ppmod}

\begin{figure}[h]
\includegraphics[width=\columnwidth]{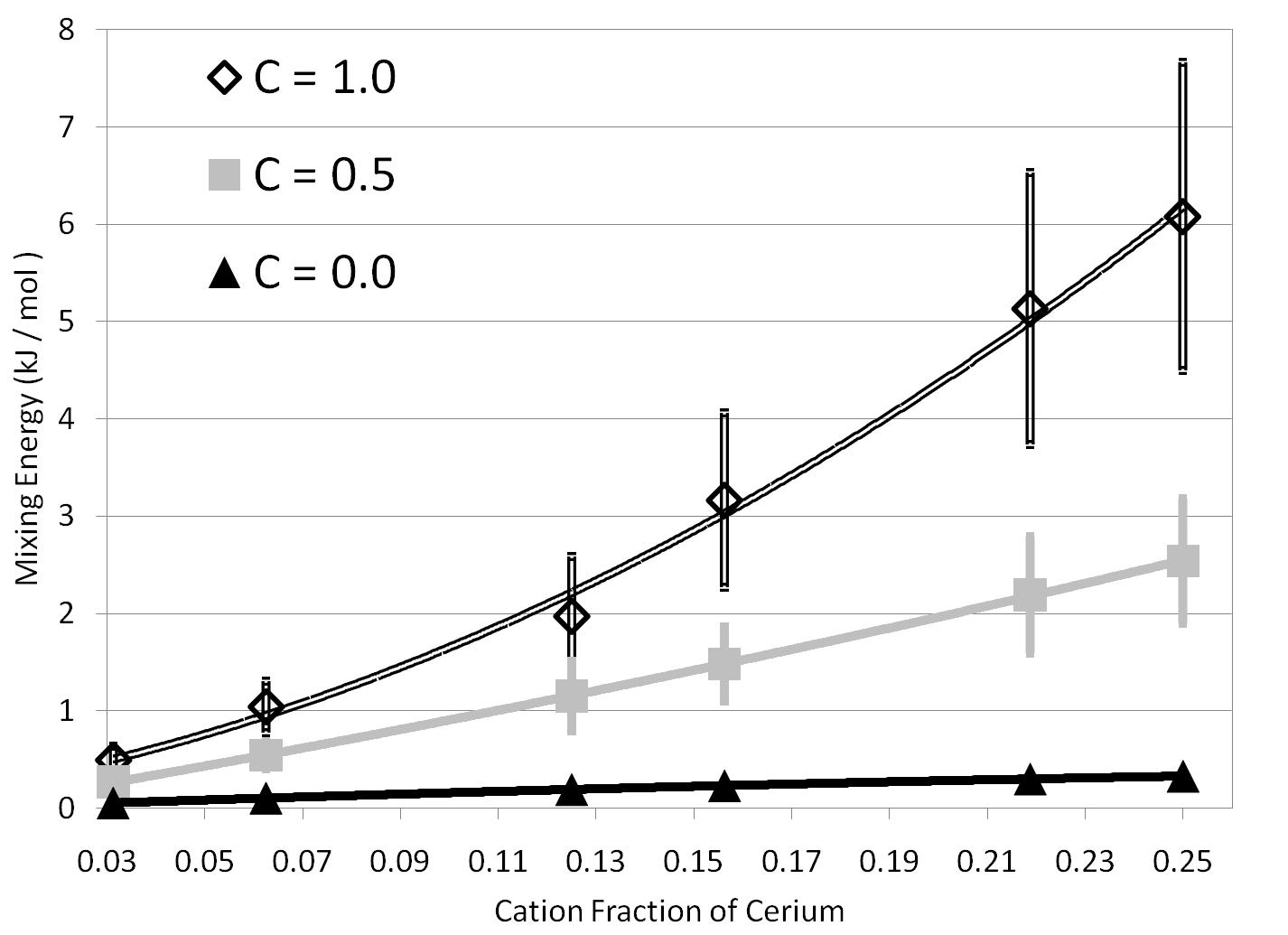}
\caption{Average mixing enthalpies obtained from ionic pair-potential calculations for structures enumerated by randomly populated supercells. The average mixing energy obtained for each composition is given by a filled symbol, and the solid lines are polynomial fits, included to highlight the trends. Error bars represent $\pm$1 standard deviation in all of the data for a given composition. ``C'' is the mole fraction of Ce which is in the Ce(III) state.}
\label{fig:pp_rand}
\end{figure}

\begin{figure*}
\includegraphics[width=2\columnwidth]{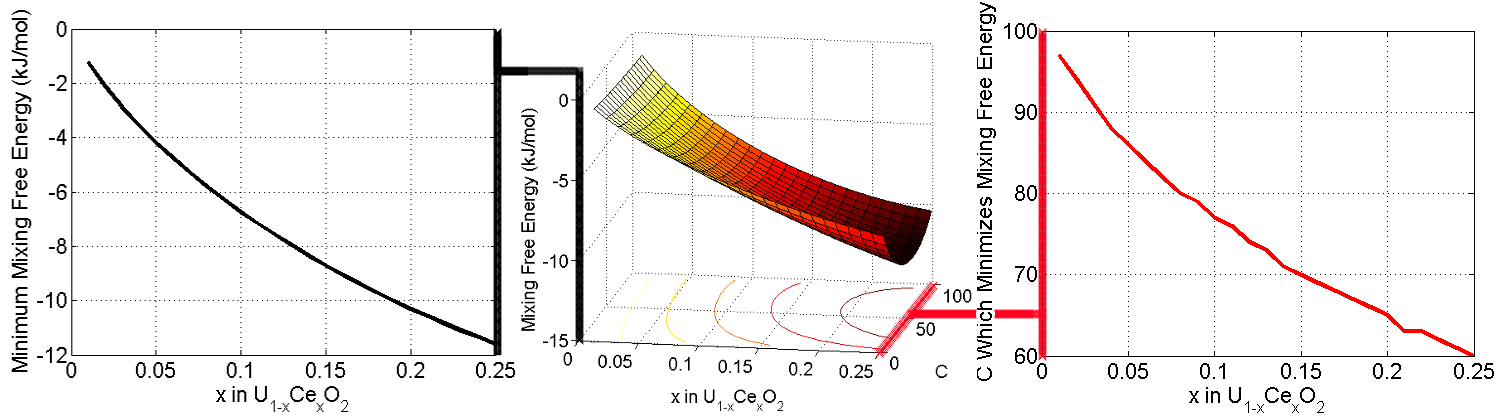}
\caption{(color online)The center graph depicts the ideal mixing free energy as a function of the two composition variables, cerium cation fraction (``x'') and the percentage of the Ce ions which are in a charge state of III (``C''). The left- and right-most panels give the values of the mixing free energy and the values of the charge transfer parameter (``C''), respectively, which correspond to the convex hull of this plane. These illustrate the lowest energy states at each total cerium composition.}
\label{fig:mixg}
\end{figure*}

The DFT+U results presented in Figs. \ref{fig:energies}-\ref{fig:mixg} have been derived for long-range-ordered compounds featuring prototypical ordering of U and Ce ions on the fcc cation sublattice.  As discussed above, stoichiometric urania-ceria mixtures observed experimentally are disordered solid solutions, lacking measurable cation order. To investigate the implications of the DFT+U results for the mixing thermodynamics of disordered solid solutions we have undertaken calculations based on classical ionic pair-potential models, as described in section \ref{ss:pp}.

Figure \ref{fig:pp_rand} plots the averaged values (symbols) and standard deviations (indicated by error bars) of the mixing energies calculated by averaging over supercell configurations with randomly generated cation and charge disorder.  The mixing energies for the ICS states are small and positive in magnitude, consistent with the DFT+U results for the ordered ICS structures in Fig. \ref{fig:energies}. With increasing degree of charge transfer, the mixing energy is seen to increase and is analogous to the DFT calculations, in which charge transfer is disfavored energetically. Finally, the increased variation in the energies of structures with high degrees of charge transfer is reflected by the larger standard deviations in the calculated mixing energies.

Although charge transfer is found to lead to a higher energy of the MCS relative to the ICS states in Figs. \ref{fig:energies} and \ref{fig:pp_rand}, the energy difference is relatively small, such that the MCS configurations are expected to be sampled for entropic reasons at the temperatures used in the synthesis and applications of urania-based fuels.  To demonstrate this point, we show in Fig. \ref{fig:mixg} the mixing free energy derived at T=1500 K, obtained by combining the calculated mixing enthalpies with an ideal mixing entropy.  Specifically, the free energy plotted in the left panel is obtained by minimizing the expression:

\begin{eqnarray}
\Delta G_{mix} = & \Delta & H_{mix}(x_{Ce(III)},x_{Ce(IV)},x_{U(IV)},x_{U(V)}) \nonumber \\
&+& RT \left[ \sum_{i} x_i ln( x_i)\right] 
\end{eqnarray}

\noindent with the mass and charge neutrality constrains:

\begin{equation}
\sum_{i} x_i = x_{Ce(III)} + x_{Ce(IV)} + x_{U(IV)} + x_{U(V)} = 1
\end{equation}
\begin{equation}
3x_{Ce(III)} + 4x_{Ce(IV)} + 4x_{U(IV)} + 5x_{U(V)} = 4
\end{equation}

\noindent where \emph{x\subs{i}} is the fractional cation lattice occupation of species \emph{i}. The free energy surface thus obtained is plotted in the middle panel of Fig. \ref{fig:mixg}.

The degree of charge-transfer is defined as $C = x_{Ce(III)}/(x_{Ce(III)}+x_{Ce(IV)})$, and its value which minimizes $\Delta$G\subs{mix} is shown in the right panel of Fig. \ref{fig:mixg}. The corresponding value of the mixing free energy is shown in the left panel of the same figure.  The results display an appreciable degree of charge transfer in dilute solutions that decreases with increasing Ce concentrations.  This behavior reflects the fact the degree of charge transfer is driven by entropic contributions that increase with increasing electronic disorder.  We note here that the entropic stabilization of mixed-charge states, suggested in the present calculations for urania-ceria solid solutions, has also been shown in recent calculations for LiFePO\subs{4} battery materials \cite{Zhou06}.

\section{Summary and Discussion}

In this work, we have examined the mixing energetics of U$_{\rm 1-x}$Ce$_{\rm x}$O$_{\rm 2}$ using DFT+U and classical pair potential methods. Results obtained for prototypical ordered structures on the cation sites result in largely structure independent, nearly ideal mixing enthalpies for systems in which U and Ce ions maintain the ideal 4+ charge state.  With the values of U\subs{eff} used in this work, DFT+U calculations show that charge transfer between uranium and cerium sites (resulting in U(V) and Ce(III) cations) leads to an increase in mixing energy.  We note that the sign and magnitude of the effect of charge transfer on the mixing energetics is sensitive to the choice of the U\subs{eff} parameters in the DFT+U calculations.  The results obtained in this work are based on values that have been found to lead to a good match (accounting for over-binding of the oxygen molecule) between calculated and measured values for relevant oxidation and reduction reactions in urania and ceria.  In the present calculations, the magnitude of the increase in $\Delta$H\subs{mix} resulting from the formation of mixed-charge states is highly sensitive to the ordering of the cations. Despite this tendency for charge transfer to destabilize the mixtures energetically, the energy difference between ICS and MCS states is small enough that charge disordering may be entropically stabilized at high temperatures (1500 K) relevant to synthesis and applications of urania solid solutions.  This point was illustrated in section \ref{ss:ppmod} with the aid of pair-potential models.

The small differences in energy between the ICS and MCS states obtained in this work suggest that the degree of charge transfer in urania-ceria solutions is likely to depend sensitively on oxygen stoichiometry. Indeed, it has been shown in XPS studies \cite{Bera09} that the concentration of Ce(III), U(V), and U(VI) are all strongly dependent on oxygen content. Notably, however, it is also possible that the degree of charge transfer could be different in the near-surface regions probed by XPS. This would provide a possible reason for why charge transfer is found in XPS studies, while no evidence for mixed charge states was obtained by XANES measurement \cite{Antonio96}. At present the most compelling evidence for charge transfer in bulk samples is that derived from magnetic susceptibility measurements \cite{Hinatsu88}.  In light of the present computational results, and the existing discrepancies in conclusions drawn from different experimental methods, multiple experimental methods applied to the same set of samples would be highly desirable.  Given that the cation interactions and oxygen stoichiometry are likely to be strongly coupled in this system, attention to characterizing composition and homogeneity would be particularly important for such future studies.

\section{Acknowledgments}
We would like to thank Alexander Thompson, Fei Zhou, and David Andersson for helpful discussions. This work was supported by the US Department of Energy, Office of Nuclear Energy, through the Nuclear Energy Research Initiative for Consortia (NERI-C) program, contract number DR-FG07-071D14893, as well as the US Department of Energy, Office of Nuclear Energy, Nuclear Energy Advanced Modeling and Simulation (NEAMS) Program.

\section{References}

\bibliography{2010_charge_ordering_aps}

\end{document}